\documentclass{llncs}

\usepackage{amsmath}
\usepackage{multirow}

\begin{document}

\title{Co-design of a particle-in-cell plasma simulation code for Intel Xeon Phi: a first look at Knights Landing}

\author{Igor~Surmin \inst{1} \and Sergey~Bastrakov \inst{1} \and Zakhar~Matveev \inst{2} \and Evgeny~Efimenko \inst{1, 3} \and Arkady~Gonoskov \inst{1, 3, 4} \and Iosif~Meyerov \inst{1} }

\institute{Lobachevsky State University of Nizhni Novgorod, Russia \and
Intel Corporation \and
Institute of Applied Physics of the Russian Academy of Sciences, Russia \and
Chalmers University of Technology, Sweden}

\maketitle

\abstract{
Three dimensional particle-in-cell laser-plasma simulation is an important area of computational physics. Solving state-of-the-art problems requires large-scale simulation on a supercomputer using specialized codes. A growing demand in computational resources inspires research in improving efficiency and co-design for supercomputers based on many-core architectures. This paper presents first performance results of the particle-in-cell plasma simulation code PICADOR on the recently introduced Knights Landing generation of Intel Xeon Phi. A straightforward rebuilding of the code yields a 2.43~x speedup compared to the previous Knights Corner generation. Further code optimization results in an additional 1.89~x speedup. The optimization performed is beneficial not only for Knights Landing, but also for high-end CPUs and Knights Corner. The optimized version achieves 100 GFLOPS double precision performance on a Knights Landing device with the speedups of 2.35~x compared to a 14-core Haswell CPU and 3.47~x compared to a 61-core Knights Corner Xeon Phi.
}

\section{Introduction}

The first supercomputer to pass the 100 PFLOPS mark (according to the TOP500 list, https://www.top500.org/) opens a new stage in the road to exascale systems. Such systems are expected to solve important problems of computational science, such as climate modeling, improving efficiency of energy sources, human brain simulation at neural level, and others. Making progress in assembling and efficient utilization of large supercomputers requires an interdisciplinary collaboration of software developers with engineers, mathematicians, physicists, chemists, and experts in other areas. The interdisciplinary principle is an important part of co-design in supercomputing. Currently, a significant share of supercomputers is based on many-core architectures, most notably GPUs and Intel Xeon Phi. Thus, it is important to co-design codes for such architectures.

In June 2016, during the ISC High Performance, the first performance results of the new Intel (R) Xeon Phi (TM) of Knights Landing (KNL) generation for solving several problems have been presented \cite{ISC,Jeffers}. New Xeon Phi devices are many-core CPUs with 60+ cores and 4 hardware threads per core, 512-bit SIMD, and 16 GB high-bandwidth MCDRAM. Compared to the previous Knights Corner (KNC) generation, the new Xeon Phi devices not only bring about 3~x improvement in the single-core performance, but also eliminate the need for a PCI Express connection, that was a major problem for the KNC coprocessors. Taking into account binary compatibility of the code between regular and KNL-generation CPUs, it is interesting to research performance of existing parallel codes on KNL as well as develop approaches to code optimization for KNL.

The studies presented in this paper are motivated with growing needs for carrying out large-scale 3D particle-in-cell simulations in several research directions of plasma physics. Performing such simulations is possible on supercomputers with specialized parallel codes. The particle-in-cell method inherently allows massively parallel processing and thus can be efficiently implemented for supercomputers. The growth of computational power accompanied with multilevel parallelization and optimization leads to gradual extension of capabilities of particle-in-cell codes, such as \cite{Fonseca,Bowers,Pukhov,Vay,Burau}, giving access to fascinating studies that have been previously impossible. 

Techniques of implementation and optimization of particle-in-cell codes for many-core architectures are rather well studied. There are several highly efficient implementations of the particle-in-cell method for GPUs, including \cite{Burau,Kong,Decyk,Glinsky}. Intel Xeon Phi is a newer platform with some specific features. Our previous work \cite{BastrakovLNCS,Surmin} was among the first attempts of implementation of the method for Xeon Phi of KNC generation, along with another study \cite{Nakashima}. The previous work showed that the KNC generation of Xeon Phi allows relatively easy porting of existing parallel codes with reasonable performance, but obtaining significant speedups over multi-core CPUs could require some additional work, most importantly in terms of vectorization.

This paper presents the first performance results of PICADOR particle-in-cell laser-plasma simulation code \cite{Surmin,Bastrakov} on Intel KNL. The code is developed by an interdisciplinary group of physicists, mathematicians, and software developers. The main contribution of this paper is performance evaluation of a plasma simulation code on high-end CPUs and Xeon Phi of KNC and KNL generations. We measure performance for a baseline, previously optimized, version of the code and show results of applying further optimization steps on CPUs and Xeon Phi.

The paper is organized as follows. Section \ref{sec_pic} briefly describes the particle-in-cell method for laser-plasma simulation. Section \ref{sec_baseline} gives performance results for the baseline version of the code without additional modification on KNL. Section \ref{sec_optimization} presents results of optimization of the code with some KNL-specific methods and some methods that yield benefit on other platforms as well. Section \ref{sec_conclusions} concludes the paper.

\section{Particle-in-cell method overview} \label{sec_pic}

The progress in utilization of supercomputers for particle-in-cell plasma simulation is of a special interest in the context of the rapid advancement of technologies of producing high-intensity laser pulses. Nowadays, high-intensity laser systems are reaching unprecedented densities of electromagnetic energy among all controllable sources available in a laboratory. Interaction of such laser pulses with various targets provides a possibility to access extreme conditions and new regimes that open up new ways towards solving important technological problems and carrying out fundamental studies, ranging from compact sources for hadron therapy to probing nonlinear properties of vacuum. Particle-in-cell simulations are known to play a key role in a wide range of related studies, because the methodology of the particle-in-cell method allows natural account for various phenomena, from target ionization at low intensities to the processes due to quantum electrodynamics at ultra-high intensities \cite{Gonoskov}. 

However, the basic stages of plasma simulation with the particle-in-cell method typically remain the most computationally demanding and challenging for optimization. The particle-in-cell method \cite{Hockney,Dawson,Birdsal} implies representing real particles of plasma with a smaller number of so-called macro-particles. Just as for real particles, the dynamics of macro-particles is governed by the relativistic equations of motion. For the sake of shortness, hereafter we write particles instead of macro-particles. 

Apart from the motion under the effect of external electromagnetic fields, the particles interact with each other through the self-generated electromagnetic field, which evolves according to the Maxwell's equations. In such a way, the electromagnetic field is affected by the particles through the current density, while the particles experience the Lorentz force due to the electromagnetic field. Both electromagnetic field and current density are defined on a discreet grid. Thus, the field is interpolated to the position of particles, while the contribution of each particle to the current density is distributed among the nearest grid nodes. The core of the particle-in-cell method consists of the following stages \cite{Birdsal}: numerical integration of Maxwell's equations, field interpolation, solving particles' equations of motion, and computing the current density created by the particles. For the rest of the paper we refer to field interpolation and solving equations of motion together as particle push, computation of current density as current deposition.

From a computational point of view, the procedures of field interpolation and current deposition concern accessing and changing two differently arranged sets of data, for the particles and for the electromagnetic field and current density values at the grid nodes. Arranging efficient calculations becomes even more complicated because of constant migration of particles between the grid cells. Thus, because of both high demands of the modern studies and the method inherent complexity, efficient implementation of the particle-in-cell method remains challenging.

\section{Baseline version} \label{sec_baseline}

PICADOR \cite{BastrakovLNCS,Surmin,Bastrakov} is a C++ code for plasma simulation based on the particle-in-cell method. The code is currently used in several research projects concerning simulation of laser-plasma interaction \cite{Gonoskov,Muraviev,Mackenroth1,Mackenroth2}. Here we briefly describe the organization of parallel processing in PICADOR, more implementation and optimization details (improving memory locality and scaling efficiency, vectorization) are given in \cite{Surmin}. The code exploits parallelism on all levels available at modern cluster systems. Distributed memory parallelism is achieved by means of spatial domain decomposition and load balancing using MPI \cite{Surmin_LB}. On the shared memory level particles are stored separately for each cell; OpenMP threads process particles in different cells in parallel. SIMD instructions are used by means of partial vectorization of loops over particles in a cell as well as manual coding of intrinsic-based implementation of some stages of the method.

Throughout this paper we use a frozen plasma benchmark problem with a $40 \times 40 \times 40$ grid, $50$ particles per cell and 1000 time steps, that can be solved on a single CPU or Xeon Phi. Apart from the single-device performance, an important aspect for utilizing supercomputers is scalability on distributed memory. These two aspects are somewhat orthogonal for the particle-in-cell method, as it allows spatial domain decomposition with communications only between neighbor domains. Scaling results of PICADOR are presented in \cite{BastrakovLNCS,Surmin,Surmin_LB}, including 90\% strong scaling efficiency on a system with 64 KNC Xeon Phi coprocessors \cite{Surmin}.

The simulations were performed in double precision using the standard cloud-in-cell particle form factor \cite{Birdsal} and the charge-conserving current deposition scheme \cite{VillasenorBuneman}. The computational experiments were performed at a node of Intel Endeavor system with Intel Xeon E5-2697 v3 (Haswell, 14 cores, 2.6 GHz, 36 MB cache), Intel Xeon Phi 7120 (KNC, 61 cores, 1.2 GHz, 30.5 MB cache), and Intel Xeon Phi 7250 (KNL, 68 cores, 1.4 GHz, 34 MB cache, 16 GB MCDRAM). Intel Xeon Phi 7250 was used in Quadrant cluster mode, all data placed in MCDRAM.

We recompiled the code, which had been previously optimized for the KNC generation of Xeon Phi, to run on KNL. Since the optimal run configuration on KNC was a single MPI process and 4 OpenMP threads per core \cite{Surmin}, first we tried running a single MPI process on KNL as well, with 1, 2, 3, and 4 threads per core. The comparison of these configurations is presented at table 1. Same as for KNC, increasing the number of threads per core is beneficial for PICADOR on KNL. For the rest of the paper we only consider configurations with 4 threads per core.

\begin{table}
\centering
\caption{Run time of the baseline version on KNL with a single MPI process and different number of OpenMP threads per core. Time is given in seconds.}
\begin{tabular}{lrrrr}
\hline\noalign{\smallskip}
\multirow{2}{*}{\textbf{Stage}} & \multicolumn{4}{c}{\textbf{\#\,threads per core}} \\ {} & 1 & 2 & 3 & 4 \\
\noalign{\smallskip}
\hline
\noalign{\smallskip}
Particle push         & 13.41 & 11.69   & 9.51  & 10.92 \\
Current deposition & 12.72 & 9.84    & 10.95 & 8.91 \\
Other                   & 0.38  & 0.41    & 0.51   & 0.44 \\
Overall                 & 26.51 & 21.94  & 20.97 & 20.27 \\
\hline
\end{tabular}
\end{table}

Table 2 presents the run time of the baseline version running a single MPI process per device with 1 OpenMP thread per core on CPU and 4 OpenMP threads per core on Xeon Phi. The KNL device outperforms both 14-core Haswell CPU and 61-core KNC, the corresponding speedups are 1.51~x and 2.43~x. Thus, just rebuilding the code for KNL with no additional optimization results in a significant speedup compared to KNC. This is not a surprising result, since the theoretical performance on KNL is about 3~x of that of KNC. In the next section we demonstrate how additional optimization of the code and choosing a better configuration of processes and threads can further improve performance on KNL.

\begin{table}
\centering
\caption{Run time of the baseline version on CPU and Xeon Phi with a single MPI process on each device. Time is given in seconds.}
\begin{tabular}{lrrr}
\hline\noalign{\smallskip}
\multirow{2}{*}{\textbf{Stage}} & \textbf{Intel Xeon} & \multicolumn{2}{c}{\textbf{Intel Xeon Phi}} \\ {} & \textbf{E5-2697 v3} & \textbf{7120 (KNC)} & \textbf{7250 (KNL)} \\
\noalign{\smallskip}
\hline
\noalign{\smallskip}
Particle push         & 18.30  & 22.69 & 10.92 \\
Current deposition & 12.02  & 25.64 & 8.91 \\
Other                   & 0.25   & 0.98   & 0.44 \\
Overall                 & 30.57  & 49.31 & 20.27 \\
\hline
\end{tabular}
\end{table}

\section{Performance analysis and optimization on Knights Landing} \label{sec_optimization}

\subsection{Choosing the optimal run configuration}

A run configuration of processes and threads can significantly influence the performance of an MPI + OpenMP code and the optimal configuration is often non-obvious \cite{Jeffers}. Thus, our first step towards increasing performance is comparison of different configurations of processes and threads. Table 3 presents comparison of different configurations of processes and threads, each running 4 threads per core. Increasing the number of processes up to 8 while keeping the overall number of threads constant yields an increase in performance, up to 1.31~x compared to the single-process configuration. A possible explanation is that in this case data layout better fits the application. However, further increasing the number of processes results in performance degradation. For the rest of the paper we use the configuration with 8 MPI processes and 34 OpenMP threads per process on KNL.

\begin{table}
\centering
\caption{Run time of several process-thread configurations on KNL for the baseline version of the code. Time is given in seconds.}
\begin{tabular}{lrrrr}
\hline\noalign{\smallskip}
\multirow{ 2}{*}{\textbf{Stage}} & \multicolumn{4}{c}{\textbf{\#processes $\times$ \#threads per process}} \\ {} & $1 \times 272$ & $2 \times 136$ & $4 \times 68$ & $8 \times 34$ \\
\noalign{\smallskip}
\hline
\noalign{\smallskip}
Particle push         & 10.92  & 9.16     & 8.51 & 8.07 \\
Current deposition & 8.91    & 7.68     & 7.60 & 7.15 \\
Other                   & 0.44    & 0.35    & 0.30 & 0.26 \\
Overall                 & 20.27   & 17.19  & 16.41 & 15.48 \\
\hline
\end{tabular}
\end{table}

\subsection{Auto-vectorization of field interpolation}

Efficient vectorization of some stages of the particle-in-cell method is not easy, particularly for field interpolation that results in an intricate memory access pattern with indirect indexing \cite{Surmin,Vincenti}. For the version of the code for CPUs and KNC we explicitly disabled compiler auto-vectorization of the corresponding loops as it resulted in some slowdown due to inefficient operations with memory. However, new instructions in AVX-512 allow some speedup on KNL due to auto-vectorization of these loops. Table 4 presents comparison of the baseline version and a version with auto-vectorization of field interpolation, which is a part of the particle push stage, on KNL. The speedup of this stage due to vectorization is 1.19~x.

\begin{table}
\centering
\caption{Run time of the baseline version and a version with auto-vectorization of field interpolation on KNL. Time is given in seconds.}
\begin{tabular}{lrr}
\hline\noalign{\smallskip}
\multirow{ 2}{*}{\textbf{Stage}} & \textbf{Baseline} & \textbf{Auto-vectorization} \\ {} & \textbf{version} & \textbf{of field interpolation} \\
\noalign{\smallskip}
\hline
\noalign{\smallskip}
Particle push         & 8.07 & 6.81 \\
Current deposition & 7.15 & 7.15 \\
Other                   & 0.26 & 0.26 \\
Overall                 & 15.48 & 14.22 \\
\hline
\end{tabular}
\end{table}

\subsection{Supercells}

A promising approach to improve performance of the particle-in-cell method on many-core architectures is grouping and processing particles by supercells, formed by several nearby cells. First introduced for GPUs \cite{Burau}, it has been recently reported to be advantageous for CPUs as well \cite{Vincenti}. The size of supercells is chosen so that particle and grid data processed during the particle push and current deposition stages fit L1 cache. 

The exact amount of data processed is implementation-specific. For PICADOR with supercells of size $S \times S \times S$ cells, the size of data used for Villasenor -- Buneman current deposition on a single core can be estimated as
\begin{multline*}
CurrentDepositionDataSize(S) = \left( 4 \textrm{ threads per core} \right) \times \\ \left({(S + 1)}^3 \textrm{ grid values} \right) \times  \left( 3 \textrm{ current components} \right) \times \left( 8 \textrm{\,Bytes per value} \right)\enspace.
\end{multline*}
Particles are processed in chunks, for each chunk results of field interpolation and some auxiliary coefficients are stored in a local array. For field interpolation and particle push with cloud-in-cell formfactor the approximate size of data is
\begin{multline*}
ParticlePushDataSize(S) = \left( 4 \textrm{ threads per core} \right) \times \\ \left( \left({(S + 2)}^3 \textrm{ grid values} \right) \times \left( 6 \textrm{ field components} \right) \times \left( 8 \textrm{\,Bytes per value} \right) + \right. \\ \left.
\left( 64 \textrm{\,Bytes per particle} + 56 \textrm{\,Bytes of auxiliary data per particle}\right) \right. \\ \left. \times \left(16 \textrm{ particles per chunk} \right) \right)\enspace.
\end{multline*}
Table 5 presents results for a single-core on KNL running 4 threads. For the sake of simplicity we consider only cubical supercells with equal number of cells for each dimension.

\setlength{\tabcolsep}{6pt}
\begin{table} \label{table_supercell_size}
\centering
\caption{Results for different supercell sizes on the single-core of KNL running 4 threads depending on the supercell size. Estimated size of data actively used while processing a supercell combined for 4 threads and run time are given.}
\begin{tabular}{llrrrrrr}
\hline\noalign{\smallskip}
\multirow{ 2}{*}{\textbf{Stage}} & {} & \multicolumn{6}{c}{\textbf{Supercell size for each dimension}} \\ {} & {} & \textbf{1} & \textbf{2} & \textbf{3} & \textbf{4} & \textbf{5} & \textbf{6} \\
\noalign{\smallskip}
\hline
\noalign{\smallskip}
Particle & Data size, KB & 12.86 & 19.97 & 31.68 & 49.15 & 73.54 & 105.98 \\
push & Time, sec. & 34.28 & 31.23 & 32.35 & 30.95 & 32.12 & 34.10 \\
\noalign{\smallskip}
\hline
\noalign{\smallskip}
Current & Data size, KB & 0.77 & 2.59 & 6.14 & 12.00 & 20.74 & 32.93 \\
deposition & Time, sec. & 40.44 & 30.69 & 30.25 & 28.85 & 28.77 & 28.00 \\
\hline
\end{tabular}
\end{table}
\setlength{\tabcolsep}{8pt}

As follows from table 5, in the single-core case the most efficient supercell size for particle push is 4. For the current deposition stage, increasing the size up to 6 leads to a steady increase in performance. However, using larger supercells results in decreasing the number of independent subproblems solved in parallel using OpenMP, which could hinder the overall performance. For example, taking into account chessboard supercell processing scheme used in PICADOR, for grid size $40 \times 40 \times 40$ of the benchmark problem and supercell size $S = 6$ there are at most (depending on implementation details, usage of ghost cells, etc.) 64 independent subproblems, that is not enough to fully saturate Xeon Phi. Thus, taking into account multi-threading, the optimal supercell size is smaller than for the single-core case. For our code the empirically best supercell size on KNL was 2, on KNC and Haswell CPU it was 2 for particle push and 4 for current deposition. Table 6 presents results of the supercell version on each device and the speedups compared to the baseline. Supercells are beneficial for all three platforms in question, with 1.21~x speedup on the Haswell CPU and 1.32~x speedup on Xeon Phi, both KNC and KNL.

\setlength{\tabcolsep}{6pt}
\begin{table} \label{table_supercells}
\centering
\caption{Run time of the version with supercells on CPU and Xeon Phi. For each device the empirically chosen best process-thread configuration was used. Time is given in seconds.}
\begin{tabular}{lrrr}
\hline\noalign{\smallskip}
\multirow{ 2}{*}{\textbf{Stage}} & \textbf{Intel Xeon} & \multicolumn{2}{c}{\textbf{Intel Xeon Phi}} \\ {} & \textbf{E5-2697 v3} & \textbf{7120 (KNC)} & \textbf{7250 (KNL)} \\
\noalign{\smallskip}
\hline
\noalign{\smallskip}
Particle push         & 16.78   & 20.93 & 5.73  \\
Current deposition & 8.22    & 15.44  & 4.77  \\
Other                   & 0.27    & 0.95   & 0.25  \\
Overall                 & 25.27  & 37.32  & 10.75  \\
\hline
\end{tabular}
\end{table}
\setlength{\tabcolsep}{8pt}

Finally, we evaluated performance of the optimized version in terms of achieved GFLOPS. We used Intel VTune Amplifier tool to count the number of arithmetic operations performed. The resulting performance is 42.5~GFLOPS on the CPU and 100~GFLOPS on the KNL device in double precision.

\section{Conclusions and Future work} \label{sec_conclusions}

This paper presents a first look at Intel Xeon Phi CPUs of Knights Landing generation for particle-in-cell plasma simulation. We use the plasma simulation code PICADOR, which has been previously ported and optimized for KNC. A simple rebuilding of the code for KNL yields a 2.43~x speedup compared to KNC in the same configuration. Choosing the optimal configuration of processes and threads for KNL and applying several techniques to improve performance leads to a 1.89~x speedup on KNL compared to the baseline version. Auto-vectorization of the field interpolation loop, which led to a slowdown on KNC, gives some benefit on KNL due to AVX-512 instruction set. Utilization of supercells gives speedup on CPU as well as on Xeon Phi. The speedup of the optimized version on a Knights Landing device is 2.35~x compared to a 14-core Haswell CPU and 3.47~x compared to a 61-core Knights Corner Xeon Phi coprocessor. The code achieves 100 GFLOP/s double precision performance on KNL.

Overall, the obtained results show that KNL is a promising platform for particle-in-cell plasma simulation. Compared to the previous-generation KNC, it opens new prospects for performance improvement. Same as for KNC, approaches to optimization are mostly shared with CPUs. It allows maintaining a single version of the code for CPUs and Xeon Phi, probably with some minor changes. Our future work includes further performance improvement, especially in terms of vectorization, for benchmark and up-to-date physical problems.
A more thorough study of performance, for example, using the cache-aware roofline model \cite{Ilic} or capabilities of Intel Advisor Roofline Analysis, is another direction of future work.

This study was supported by the RFBR, project No. 15-37-21015. I.~Surmin, S.~Bastrakov and I.~Meyerov are grateful to Intel Corporation for 
access to the system used for performing computational experiments presented in this paper.

\end{document}